\documentclass[
 reprint, amsmath,amssymb, amsfonts,aps,prl,
 ]{revtex4-2}
\usepackage{dcolumn}
\usepackage{bm}

\usepackage[utf8]{inputenc}
\usepackage[T1]{fontenc}
\usepackage{mathptmx}
\usepackage{hyperref}
\usepackage{mathrsfs}
\usepackage[dvipsnames]{xcolor}
\usepackage{subcaption}
\usepackage[normalem]{ulem}
\usepackage{float}
\usepackage{siunitx}

\usepackage{textcomp}
\usepackage{flafter}
\usepackage{graphicx}
\usepackage{dcolumn}

\makeatletter

\def\bibstyle#1{}
\def\bibdata#1{}
\makeatother

\begin{document}
\title{
 Phase Transitions as Emergent Geometric Phenomena\\
 A Deterministic Entropy Evolution Law
}
\author{Loris Di Cairano}
\email{l.di.cairano.92@gmail.com, loris.dicairano@uni.lu}

\affiliation{Department of Physics and Materials Science, University of Luxembourg, L-1511 Luxembourg City, Luxembourg}


\begin{abstract}
We show that thermodynamics can be formulated naturally from the intrinsic geometry of phase space alone—without postulating an ensemble, which instead emerges from the geometric structure itself. Within this formulation, phase transitions are encoded in the geometry of constant-energy manifold: entropy and its derivatives follow from a deterministic equation whose source is built from curvature invariants. As energy increases, geometric transformations in energy-manifold structure drive thermodynamic responses and characterize criticality. 
We validate this framework through explicit analysis of paradigmatic systems—the 1D XY mean-field model and 2D $\phi^4$ theory—showing that geometric transformations in energy-manifold structure characterize criticality quantitatively. The framework applies universally to long-range interacting systems and in ensemble-inequivalence regimes.
\end{abstract}

\maketitle

In physics, predictable theories are usually defined by a coherent triad: state variables, equations of motion, and intrinsic sources. State variables describe the system's configuration; equations of motion govern their evolution. In classical and quantum mechanics, this structure is self-contained: forces in classical mechanics and Hamiltonian operators in quantum mechanics provide intrinsic, deterministic dynamics.
Statistical mechanics occupies a singular position. Its scope is to bridge the deterministic microscopic world with macroscopic thermodynamics. Fundamental state variables—entropy, for instance—do not obey intrinsic evolution laws but rather condense the global properties of microscopic configurations. This is not a defect but the essential feature of thermodynamics. It stems naturally from the symplectic structure $(\Lambda,H,\omega)$ of phase space, which furnishes Hamilton's equations and the Liouville measure. However, this structure alone does not determine which microstates are accessible at a given macroscopic configuration. Bridging this gap has required postulating statistical ensembles $\rho_H$, making the theory predictive through integration rather than by solving the equations of motion. Although this asymmetry between statistical mechanics and other theories does not constitute a scientific crisis, it motivates a foundational question: 

\noindent
\emph{Is the ensemble-based structure of statistical mechanics the only one that predicts experimentally verifiable thermodynamics?} If not, \textit{can statistical mechanics admit a deterministic evolution law for entropy intrinsic to the theory itself?}

In this Letter, we show that thermodynamics emerges naturally from the intrinsic geometry of phase space. We formulate the theory axiomatically from $(\Lambda,H,\omega,\eta)$, where the metric tensor is introduced alongside the Hamiltonian and the symplectic form. Crucially, the ensemble is not postulated but emerges necessarily from geometry alone. Uncovering a sort of thermodynamic covariance within this formalism, namely an equivalence class of metrics all inducing the same thermodynamic description, we show that the microcanonical measure coincides exactly with the area of constant-energy hypersurfaces. Then, we derive an exact deterministic evolution law for entropy—the Entropy Flow Equation (EFE)—whose source is built entirely from the curvature invariants of energy hypersurfaces. This geometric identification reveals that phase transitions (PTs) correspond to geometric changes in the energy hypersurfaces.

The notion that geometry encodes thermodynamic information has a long history, explored through both geometric and topological perspectives. Rugh showed that inverse temperature is connected to Gauss curvature~\cite{rugh1997dynamical,rugh1998geometric,rugh2001microthermodynamic}. Pettini, Franzosi, and collaborators developed topological approaches linking topological changes of the energy-manifold to PTs~\cite{franzosi2000topology,franzosi2007topology,CASETTI2000237,franzosi1999topological}, stimulating an debate on the relation between topology and thermodynamic~\cite{kastner2011phase,casetti2009kinetic,casetti2006nonanalyticities,mehta2012energy,baroni2024simplified,angelani2003topological,angelani2005topology,baroni2005topological,kastner2008phase,risau2005topology}

Later work established direct identifications between geometric properties and microcanonical observables~\cite{gori2018topological,gori_configurational,di2022geometrictheory,di2021topology}. However, these contributions treat geometry as an additional structure imposed on a prescribed ensemble.
Our work reveals that thermodynamics arises from phase-space geometry itself, without ensemble assumptions. The framework is intrinsically microcanonical, applying naturally to finite systems, long-range interactions, and ensemble-inequivalent regimes where conventional methods—Lee–Yang theory, renormalization group—may be ill-defined~\cite{campa2009statistical,bouchet2010thermodynamics,dunkel2006phase}. We solve the EFE exactly and validate it against independent thermodynamic methods on paradigmatic systems—the long-range 1D XY mean-field model and the 2D $\phi^4$ model—demonstrating that geometric transformations in energy-manifold curvature quantitatively characterize PTs. This geometric foundation unifies the treatment of criticality across disparate physical regimes.

Classical statistical mechanics is grounded in the Hamiltonian quadruple \((\Lambda,H,\omega,\rho(H))\).
Here \(\Lambda\) is the phase space with coordinates \(\bm x=(\bm p,\bm q)\) and dimension $2n$, \(H\) is the Hamiltonian, \(\omega=dp_i\wedge dq^i\) is the canonical symplectic form, and \(\rho(H)\) is the statistical weight. We will focus on the microcanonical statistical ensemble.
The dynamics is generated by the Hamiltonian vector field \(\bm X_H\) via~\cite{arnol2013mathematical}
\[
\iota_{X_H}\omega=dH,
\]
so that trajectories lie on energy-manifold \(\Sigma_E:=H^{-1}(E)\) and energy is conserved,
\[
\dot H=dH(\bm X_H)=\omega(\bm X_H,\bm X_H)=0.
\]
While \(\omega\) furnishes the Liouville measure \(d\mu_\Lambda=\omega^{\wedge n}/n!=dq^1\!\cdots dq^N\,dp_1\!\cdots dp_N\) (for \(\dim\Lambda=2n\)), it does \emph{not} provide a canonical measure on \(\Sigma_E\) because of three structural obstructions; see Sec.~A in Supplemental Materials (SM): \textit{(i)} dimensional mismatch (energy shells have odd dimension \(2n\!-\!1\)); \textit{(ii)} non-uniqueness of a transverse projection; \textit{(iii)} coisotropy/degeneracy of \(\omega|_{\Sigma_E}\).
In the standard framework, this gap is bridged by \emph{postulating} an ensemble; in the microcanonical case, one assumes
\[
    d\mu_\rho=\rho(H)\,d\mu_\Lambda,\qquad \rho(H)\propto\delta(H-E),
\]
which makes the theory predictive by connecting the microscopic Hamiltonian flow to macroscopic thermodynamics.

In this work, we take a step back and construct the bridge geometrically.
Retaining \((\Lambda,H,\omega)\), we introduce a natural metric tensor, the Euclidean one: \(\eta=\delta^{ij}dp_i\!\otimes dp_j+\delta_{ij}dq^i\!\otimes dq^j\) on \(\Lambda\).
The metric tensor allows the connection \(\eta(\nabla^\eta H,Y)=dH(Y)\), due to the induced isomorphism between covectors and vectors. As a consequence, the energy conservation law is converted into the orthogonality relation
\[
    \eta(\nabla^\eta H,\bm X_H)=dH(\bm X_H)=\omega(\bm X_H,\bm X_H)=0,
\]
which identifies, at each \(\bm x\in\Lambda\),
\[
    T_{\bm x}\Lambda=\mathrm{span}\{\nabla_\eta H(\bm x)\}\oplus T_{\bm x}\Sigma_E,
\]
with \(T_{\bm x}\Sigma_E=\{Y\in T_{\bm x}\Lambda:\eta(Y,\nabla^\eta H)=0\}\). In other words, $\Lambda$ decomposes into two orthogonal directions: one is identified by $\bm X_H$, which always lies on $\Sigma_E$ and generates the Hamiltonian flow; the second is identified by $\nabla_\eta H$ and generates the motion of points on an energy hypersurface to others. In general, $\nabla_\eta H$ moves points on the same $\Sigma_E$ to different shells (see Sec.~B in SM). The only way to generate an energy motion that advances all points of \(\Sigma_E\) to the \text{same} nearby shell \(\Sigma_{E+\epsilon}\) consists of introducing an \emph{energy clock}, i.e., a \(\bm\zeta\in\mathrm{span}\{\nabla^\eta H\}\) such that \cite{hirsch2012differential} (see Sec.~B in SM)
\[
    dH(\bm\zeta)=1\quad\Longrightarrow\quad \bm\zeta=\frac{\nabla^\eta H}{\|\nabla^\eta H\|^2}.
\]
This structure is not imposed by geometry but is clearly dictated by physics; that is, the presence of a Hamiltonian function which selects the right way to move from a value of energy to another.

This vector field generates the \emph{thermodynamic dynamics}: the flow \(\Phi^{\mathrm{diff}}_{\epsilon}:\Sigma_E\to\Sigma_{E+\epsilon}\) defined by
\[
    \frac{d}{d\epsilon}\,\Phi^{\mathrm{diff}}_{\epsilon}(\bm x)=\bm\zeta(\bm x(\epsilon)),\qquad H(\bm x(\epsilon))=E+\epsilon,
\]
which yields diffeomorphisms between neighboring energy hypersurfaces. This motion naturally defines the adapted coordinates \((E,y^\alpha)\) with \(E=H(\bm x)\) and \(y^\alpha\) coordinates on \(\Lambda\). In these coordinates, the metric is block diagonal (see Sec.~C in the SM for more details):
\[
    \eta=\frac{dE\otimes dE}{\|\nabla^\eta H\|_\eta^{2}}+(\sigma^\eta_E)_{\alpha\beta}\,dy^\alpha\!\otimes dy^\beta,
\]
where \(\sigma^\eta_E\) is the metric induced on \(\Sigma_E\).
The natural measure induced on \(\Sigma_E\) is
\[
\begin{split}
    \begin{split}
    d\mu^\eta_E&=d\mu_{\Lambda}\Big|_{\Sigma_E}:=\frac{d\sigma^\eta_E}{\|\nabla^\eta H\|_\eta},
    \end{split}
\end{split}
\]
with $d\sigma^\eta_E=\sqrt{\det\sigma^\eta_E}\,dy^1\!\cdots dy^{2n-1}$ and this yields the \emph{(geometric) density of states}
\[
    \begin{split}
    \Omega_\eta(E):=\int_{\Sigma_E}d\mu_{E}^{\eta}=\int_{\Sigma_E}\frac{d\sigma^\eta_E}{\|\nabla^\eta H\|_\eta}
    =\int_{\Lambda}\delta\!\big(H(\bm x)-E\big)\,d\mu_\Lambda.
    \end{split}
\]
where the coarea formula has been used~\cite{federer2014geometric}. The last integral on the right coincides with the Boltzmann density of state $\Omega_B(E)$. This result embodies a conceptual inversion of the standard statistical mechanics framework. Ordinarily, one postulates an ensemble $\rho_H$ first, and then—if desired—adds geometric structure and passes from right to left. Here, we reverse the logical order: the metric tensor is introduced axiomatically alongside the fundamental Hamiltonian and symplectic structure, and the microcanonical ensemble emerges \emph{necessarily} as a purely geometric consequence. The metric supplies the missing transverse geometry, internalizing the ensemble not as an external postulate but as an intrinsic property of phase space itself.

The only physical constraint imposed by the symplectic structure on the metric tensor is the conservation of Liouville measure,
$\mathscr{L}_{\bm{X}_H}(d\mu_{\Lambda})=0$, and then $\mathscr{L}_{\bm{X}_H}d\mu^\eta_{E}=0$. This constraint fixes the form of the measure but not the metric tensor itself. Hence, there exists an \emph{equivalence class} $[\eta]$ of metrics such that if $g$ is such that 
\[
    d\mu^g_E=\frac{d\sigma^g_E}{\|\nabla^g H\|_g}\overset{\text{area}}{\simeq}\frac{d\sigma^\eta_E}{\|\nabla^\eta H\|_\eta}=d\mu^\eta_E\,,
\]
then $g\in[\eta]$ and it is thermodynamically equivalent to $\eta$. This is a \textit{geometric gauge symmetry of thermodynamics} that we formulate as follows:\\
\textbf{Thermodynamic covariance}. The thermodynamic content of a Hamiltonian system is independent of the geometric representation of phase space, provided that the microcanonical weight is preserved.

The equivalence class $[\eta] = \{g : g \sim \eta\}$ consists of all metrics that induce the same microcanonical measure $\Omega_\eta$. Therefore, all microcanonical observables, namely, $S(E)$ and its energy derivatives, depend only on the equivalence class $[\eta]$.

\begin{figure*}
    \centering
    \includegraphics[width=0.8\linewidth]{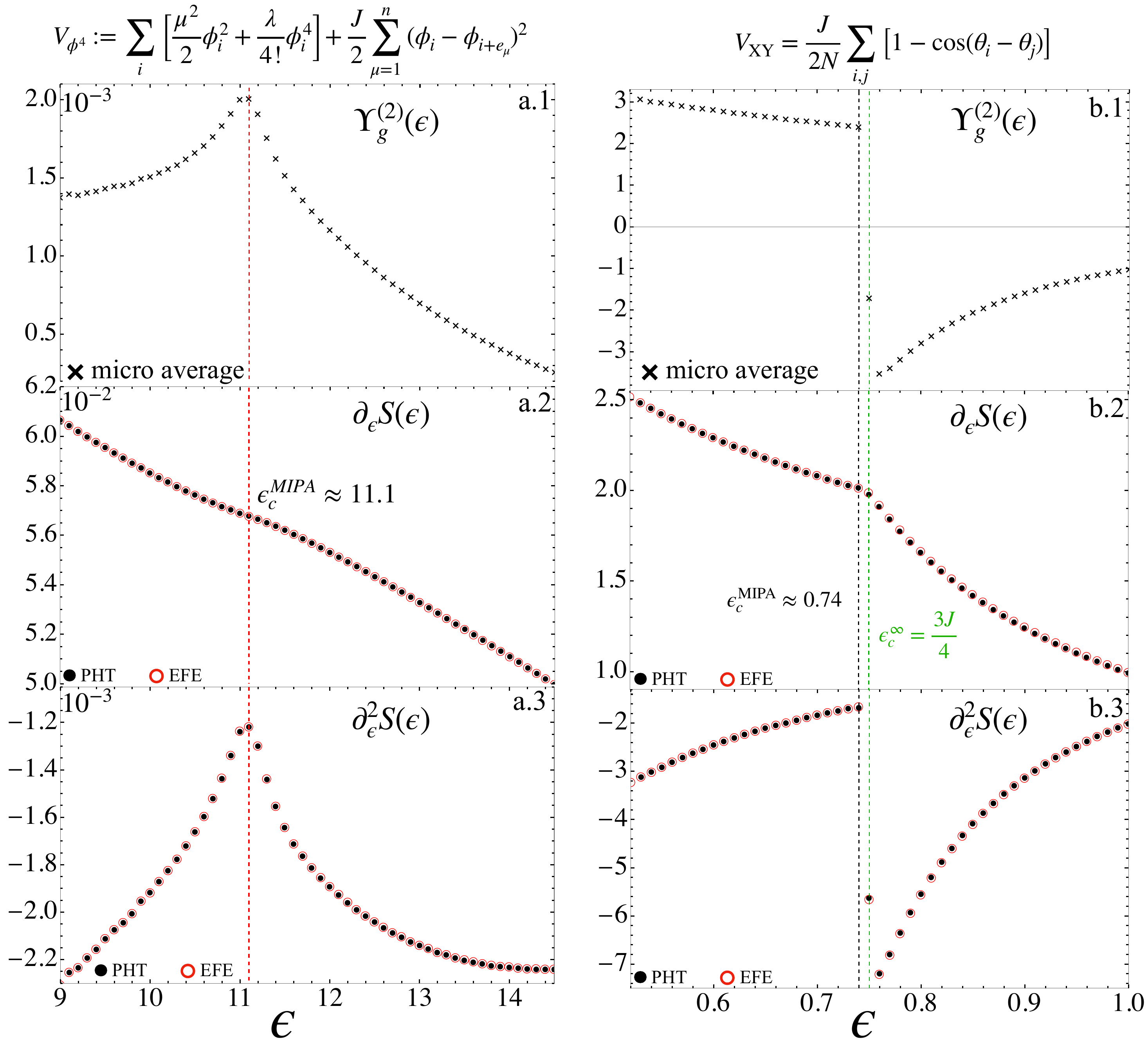}

    \caption{\textbf{Comparison of entropy derivatives obtained as the solution of the entropy flow equation and through thermodynamic methods}. Empty red circles are associated with the solutions (derivatives of entropy) of the Entropy Flow Equation (EFE). Black disks are the numerical estimations of entropy derivatives obtained from  microcanonical simulations with Pearson-Halicioglu-Tiller method \cite{pearson1985laplace}. The geometric function $\Upsilon^{(2)}_g$ represented by black crosses is computed through numerical simulation according with Eq.~(S-27) in SM. In panels a.1-a.3, we report results for the 2D $\phi^4$-model with nearest interactions and coupling parameters $\lambda=3/5$, $\mu=\sqrt{2}$ and $J=1$ and size $N=256^2$. The vertical red dashed line indicates the critical energy detected by MIPA and in agreement with the literature. Panels b.1-b.3 are associated with the comparison of entropy derivatives in the 1D XY mean-field model with coupling parameter $J=1$ and size $N=40000$. Notice that the infinite-size critical energy $\epsilon^\infty_c=3J/4$ is closed to the finite-size critical energy $\epsilon_c^{\text{MIPA}}=0.74$ estimated by MIPA. See SM for more details. }
    \label{fig:solution-riccati-xy}
\end{figure*}

We are then allowed to choose the most convenient element $g$ within the class $[\eta]$. Such a selection, corresponding to fix the gauge, follows by the request that the generator $\bm\xi$ of the diffeomorphisms $\Phi^{\rm diff}_{\bm\xi}:\Sigma_E\mapsto\Sigma_{E+\delta E}$ has unit length, $\|\bm\xi\|_g=1$. A straightforward calculation (see Sec.~D in SM) shows that the unit-length implies $\bm\xi=\nabla^g H$ and $\text{area}_g=\Omega_\eta$. We call this metric the \emph{unit-normal gauge}. This leads to the following results:\\
\textbf{Geometric Microcanonical Equivalence}. Within the thermodynamically equivalent metric class, it is possible to find a geometric representation $g$ such that counting accessible states is equivalent to measuring the occupied geometric area \cite{gori2022topological}.
Entropy follows as an intrinsic geometric property:
\begin{equation}\label{def:geometric-entropy}
    S_g(E):=\ln\mathrm{area}^{g}(E)=\ln\Omega_B(E)=S_B(E)\,,
\end{equation}

This result suggests that geometry itself dictates the full thermodynamic behavior.
Intuitively, the thermodynamic properties of a physical system and the emergence of PTs are encoded within the geometry of the energy hypersurfaces and detected through the energy flow $\Phi^{\rm diff}_{\bm\xi}$.

We show how an evolution flow with an intrinsic geometric source naturally emerges. Differentiating Eq.~\eqref{def:geometric-entropy} twice:
\begin{align}
        \partial_{E}S_{g}(E)&=\frac{\partial_{E}\text{area}^{g}(E)}{\text{area}^{g}(E)},\label{eqn:first_derivatives_entropy}\\
        \partial_{E}^{2}S_{g}(E)&=\frac{\partial_{E}^{2}\text{area}^{g}(E)}{\text{area}^{g}(E)}-\left(\frac{\partial_{E}\text{area}^{g}(E)}{\text{area}^{g}(E)}\right)^{2}.\label{eqn:second_derivatives_entropy}
\end{align}

A recursive structure naturally appears, $\Upsilon^{(k)}(E)=\partial_E^k \text{area}^{g}(E)/\text{area}^{g}(E)$, whose members can be expressed in terms of geometric invariants. The first variation of volume involves the trace of the Weingarten operator since $\mathscr{L}_{\bm\xi} d\mu_g=(\mathrm{div}_g \bm\xi)\,d\mu_g$ with 
\[
    \mathrm{div}_g\bm\xi=\mathrm{Tr}^g[W_g]=\Delta_gH\,.
\]
Note that this can be written in terms of the Laplacian, gradient, and Hessian associated with the metric $\eta$, and it reads
\begin{equation}\label{def:trace-weingarten}
\begin{split}
        \text{Tr}^g[W_g]=\frac{\Delta_\eta H}{\|\nabla_\eta H\|^2_{\eta}}-2\frac{ \langle\nabla_\eta H,{\rm Hess}_\eta\,H\,\nabla_\eta H\rangle_{\eta}}{\|\nabla_\eta H\|^4_{\eta}}\;.
\end{split}
\end{equation}
Therefore:
\begin{align}\label{eq:first-variation}
    \partial_ES_g(E)\equiv\Upsilon^{(1)}(E)=\int_{\Sigma_E}\frac{\mathscr{L}_{\bm\xi}(d\mu_g)}{\text{area}^g(E)}=\int_{\Sigma_E} \text{Tr}^g[W_g]~d\rho_g\,,
\end{align}
where $d\rho_g:=d\mu^g_E/\text{area}^{g}(E)$. The microcanonical inverse temperature $\beta(E)\equiv\partial_ES_g(E)$ coincides with the geometric measure of the Weingarten operator, i.e., 
\[
    \beta(E)=\langle\text{Tr}^gW_g\rangle_{\Sigma_E}\,.
\]
This is the functional found by Rugh~\cite{rugh1997dynamical,rugh2001microthermodynamic}, expressed here in a fully geometric way and already established in Ref.~\cite{gori_configurational}.

Taking the second variation, we have $\mathscr{L}_{\bm \xi}\!\left(\text{Tr}^g[W_{g}]\,d\mu_g\right)=\text{Tr}^g[\nabla_{\bm\xi}W_{g}]+\text{Tr}^g[W_{g}]^2d\mu_g$. Then, using~\cite{gromov2019four}:
\[
    \text{Tr}^g[\nabla_{\bm{\xi}}W_g]=-\text{Tr}^g[W_g^{2}]-Ric(\bm{\xi},\bm{\xi})\,,
\]
where $Ric$ is the Ricci curvature along $\bm{\xi}$ and the Gauss-Codazzi equation $2\,Ric(\bm{\xi},\bm{\xi})=R^g_{\Lambda}-R^g_{\Sigma_{E}}-\text{Tr}^g[W_g^{2}]+\text{Tr}^g[W_g]^{2}$, we obtain
\begin{equation}\label{def:second-variation-volume}
    \Upsilon^{(2)}_g(E)=\frac{1}{2}\!\int_{\Sigma_{E}}\!\!
    \Big\{\text{Tr}^g[W_g]^{2}-\text{Tr}^g[W_g^{2}]
    +R^{g}_{\Sigma_{E}}-R^{g}_{\Lambda}\Big\}d\rho_g,
\end{equation}
where $R^{g}_{\Sigma_{E}}$ and $R^{g}_{\Lambda}$ are, respectively, the scalar curvatures of $\Sigma_{E}$ and of the phase space \cite{petersen2006riemannian}. The explicit expression for $\Upsilon^{(2)}_g$ computed in simulations is given in Eq.~(S-27) in the SM.  
Combining Eqs.~\eqref{eqn:first_derivatives_entropy}–\eqref{eqn:second_derivatives_entropy}, we obtain
\begin{equation}\label{eqn:EFE}
    \partial_{E}^{2}S_{g}(E)+[\partial_{E}S_{g}(E)]^{2}=\Upsilon^{(2)}_g(E),
\end{equation}
a Riccati-type differential law for the entropy, hereafter called the \emph{Entropy Flow Equation} (EFE).  
Equation~\eqref{eqn:EFE} reveals that entropy is governed by a deterministic geometric flow: once the initial conditions are specified, its evolution with energy is uniquely determined by the curvature invariants of $\Sigma_E$.  

This marks a change in perspective—entropy, traditionally introduced as a statistical construct, becomes a dynamical variable whose evolution is dictated by geometry itself.  
Variations of $\Upsilon^{(2)}_g$, which encode changes in the curvature of $\Sigma_E$, translate directly into thermodynamic responses. In this sense, geometry encodes thermodynamics.

To validate the predictive power of the EFE, we applied it to two paradigmatic models:  
the 2D $\phi^4$ model with nearest-neighbor interactions and the long-range 1D XY mean-field model. All technical details about simulations are reported in Sec.~H of the SM.
In both cases, we evaluate the geometric function $\Upsilon^{(2)}(E)$ (Eq.~(S-27) in the SM) through microcanonical simulations over a dense grid of energies, as reported in panels a.1 and b.1 of Fig.~\ref{fig:solution-riccati-xy}).  
The resulting $\Upsilon^{(2)}(E)$ was then inserted into Eq.~\eqref{eqn:EFE}, which we solved numerically using a high-order Runge–Kutta integrator, yielding the entropy derivatives $\partial_E S_g$ and $\partial_E^2 S_g$. These observables are compared with those obtained through the Pearson-Halicioglu-Tiller (PHT) method~\cite{pearson1985laplace}, which is a fully thermodynamic approach and geometry-independent.

In Fig.~\ref{fig:solution-riccati-xy}, panels a.2-a.3 for the $\phi^4$ model and b.2-b.3 for the XY mean-field model, we report the comparison between geometric and thermodynamic entropy derivatives. The excellent quantitative agreement between them demonstrates that the EFE faithfully reproduces both the qualitative and quantitative energy dependence of entropy and its derivatives. Geometry contains information about thermodynamics and PTs.

Both systems undergo a second-order PTs, as is well-known by the literature. Here, we show that the microcanonical inflection point analysis (MIPA)~\cite{qi2018classification} is able to coherently detect the PT even at a finite-size. A deeper analysis with increasing system sizes, especially for the 1D XY mean-field model, is reported in Ref.~\cite{PRE-loris}. 

For the $\phi^4$ model, $\partial_E S_g$ displays an inflection point (panel a.3 and $\partial_E^2 S_g$ a negative-valued maximum at $\epsilon_c\simeq 11.1$ (panel a.2, in agreement with MIPA for a second-order PT. The critical point is also in agreement with the literature \cite{mehta2012energy,kastner2011phase,bel2020geometrical}.  
For the XY mean-field model, the same signatures appear at $\epsilon_c^{\mathrm{MIPA}}\!\simeq\!0.73$; see panels b.3 and b.2. This value corresponds to the finite-size critical energy, while in the thermodynamic limit, we have $\epsilon_c^{\infty}\!=\!3J/4$ \cite{antoni1995clustering,campa2014physics}. We will show below, through an analytical approach within the geometric framework, how to determine the infinite-size critical energy. Notably, the curvature of $\partial_E^2 S_g$ changes more sharply in the XY case, reflecting the stronger collective character of its transition.

Apart from the exact, that is, the non-perturbative solution provided by solving the EFE~\eqref{eqn:EFE}, it is crucial to determine the microscopic geometric mechanism that entails a PT. To this scope, we perform an explicit analytical expansion of $\mathrm{Tr}\,W_g$ in the paradigmatic 1D XY mean-field model, whose Hamiltonian is
\begin{equation}
    H_{\text{XY}}(\theta,p)=\sum_{i=1}^N \frac{p_i^2}{2}+\frac{J}{2N}\sum_{i,j}\bigl[1-\cos(\theta_i-\theta_j)\bigr].
\end{equation}  
We remark that $\mathrm{Tr}\,W_g$ plays a central thermodynamic role, being related to the inverse temperature. 
 
The strategy consists of expanding $\mathrm{Tr}\,W_g$ in powers of $M$, i.e.,
\[
    \mathrm{Tr}[W_g]=C^{\infty}_0(\epsilon)+C^{\infty}_2(\epsilon)\,M^2+O(M^4),
\]
where $M$ is the magnetization (order parameter), defined by \( \bm M=N^{-1}\sum_i(\cos\theta_i,\sin\theta_i)=(M\cos\phi,M\sin\phi)\), and 
where $C_0$ and $C_2$ are $\epsilon$-dependent coefficients that survive in the thermodynamic limit.  
We will find that the critical point $\epsilon_c$ corresponds to the energy density at which the concavity of the energy manifolds changes sign. All details are reported in the SM.

We begin by noting that $\mathrm{Tr}\,W_g$ in Eq.~\eqref{def:trace-weingarten} can be exactly decomposed into three explicit blocks associated with momenta, $\kappa_p$, diagonal angular curvature, $\kappa^{\rm d}$, and angular interactions, $\kappa^{\rm int}$:
\begin{equation}
    \text{Tr}[ W_g]=\sum_i\bigg(\kappa_{p_i}+\kappa^{\text{d}}_{\theta_i}+\sum_{j\neq i}\kappa^{\mathrm{int}}_{ij}\bigg)
\end{equation}
and, setting $\Theta_i=\theta_i-\phi$, they read
\begin{align}
    \label{def:norm-grad}
    &K:=\sum_{i=1}^N p_i^2,\qquad\qquad G:=\|\nabla H\|^2=K+J^2 M^2 S_2\\
    \label{def:block-P}
    &\sum_i\kappa_{p_i}=\frac{N}{G}-\frac{2K}{G^2},\qquad\qquad\qquad S_2=\sum_i \sin^2\Theta_i\\
\label{def:block-D}
    &\sum_i\kappa^{\text{d}}_{\theta_i}=\frac{JN M^2}{G}\!-\!\frac{2J^3M^3S_{21}}{G^2},\quad S_{21}=\sum_{i=1}^N \sin^2\Theta_i\,\cos\Theta_i,\\
\label{def:block-I}
    &\sum_{i\neq j}\kappa^{\mathrm{int}}_{ij}=\frac{2J^3 M^2}{N\,G^2}S_{\rm int},\;\;\; S_{\mathrm{int}}=\sum_{i\neq j}\sin\Theta_i\sin\Theta_j\cos(\theta_i-\theta_j).
\end{align} 
The next step is to expand each building block above up to order $O(M^2)$ and retain only terms of order $O(1)$ in $N$, since they are the only ones that survive in the thermodynamic limit. The angular functions are written as follows (see Sec.~F in SM):
\begin{align}    \label{def:S_expansion}
    S_2&\simeq N\Big(\frac12-\frac14 M^2\Big),\quad
    S_{21}\simeq \frac{MN}{4} ,\quad
    S_{\mathrm{int}}\simeq\frac{N^2}{4}.
\end{align}
Then, introducing the (twice) kinetic density $c:=K/N$, the norm squared in Eq.~\eqref{def:norm-grad} becomes \( G\simeq N(c+J^2M^2/2)\). This expansion is then plugged into the denominator of $\kappa_{p_i}$, $\kappa^{\rm d}_{\theta_i}$, and $\kappa^{\rm int}_{ij}$, together with Eq.~\eqref{def:S_expansion}. Then, we use the binomial series $(1+x)^{\alpha}\simeq1+\alpha x+O(x^{2})$ to expand $G$ in the denominators. In $\kappa_{p_i}$, only the first term $N/G$ is leading in $M^2$ and of order $O(1)$ in $N$; it reads 
\[
\begin{split}
    \sum_i\kappa_{p_i}\simeq
   \frac{1}{c}-\frac{J^2}{2c^2}M^2+O(M^4)\,,
\end{split}
\]
since $2K/G^2\simeq 2(1-2J^2M^2/c+O(M^4))/Nc$ is therefore negligible in the limit $N\to\infty$, as it is of order $O\left(1/N\right)\left(1+O(M^2)\right)$. Similarly, in $\kappa^{\rm d}_{\theta_i}$, only the first term, namely, $JNM^2/G$, survives, and it reads 
\[
    \sum_i\kappa^{\rm d}_{\theta_i}\simeq \frac{JM^2}{c}\;.
\]
In fact, the second term gives \( 2J^3M^3S_{12}/G^2=J^3M^4/2c^2N+O(M^6/N)\) since $S_{21}=O(N M)$ and is thus negligible at order $M^2$ in the thermodynamic limit.
Finally, the $\kappa^{\rm int}_{ij}$-term is entirely sub-leading at order $M^2$ and vanishes in the large–$N$ limit; indeed
$\sum_i\kappa^{\rm int}_{ij}\simeq J^3 M^2/2N c^2\simeq O(M^2/N)$.

Collecting all terms, the trace reads:
\begin{equation}
    \mathrm{Tr}[W_{g}]^{\infty}\approx \frac{1}{c}+\Big(\frac{J}{c}-\frac{J^2}{2c^2}\Big)\,M^2 ,
\end{equation}
and expressing it in terms of the energy density $\varepsilon$ yields
\begin{equation}
    C^{\infty}_2(\varepsilon)=\frac{J}{2\varepsilon-J}-\frac{J^2}{2(2\varepsilon-J)^2}
    =\frac{2J}{(2\varepsilon-J)^2}\!\left(\varepsilon-\frac{3J}{4}\right)\!.
\end{equation}
This analytical result makes the geometric nature of the transition explicit. The change of sign of $\mathrm{Tr}\,W_g$ at $\varepsilon_c=3J/4$ corresponds to a reversal of curvature in the energy manifolds—an instability of the $M=0$ branch where the geometry itself selects the ordered phase. The agreement with the known thermodynamic critical point confirms that the curvature of the energy manifolds encodes the full thermodynamic response. The 1D XY mean-field model, therefore, provides a transparent illustration of how a PT emerges as a purely geometric change.

In summary, we have shown that thermodynamics can be formulated in a fully geometric fashion in terms of $(\Lambda,H,\omega,\eta)$ without postulating a priori the concept of (microcanonical) ensemble; rather, the latter arises naturally, and entropy becomes a geometric quantity whose evolution is governed by the deterministic entropy flow equation. This equation connects the derivatives of entropy to the curvature invariants of the energy manifold, thus providing a self-contained relation between dynamics and thermodynamics. The explicit analysis of the 1D XY mean-field model demonstrates that PTs correspond to geometric changes in the curvature of the energy manifolds---which reproduce known critical points.  

Combining the EFE with Bachmann’s microcanonical inflection-point analysis (MIPA)~\cite{qi2018classification} we can establish a direct geometric foundation for PTs: each critical behavior classified by MIPA corresponds to a specific geometric profile of the source term $\Upsilon^{(2)}_g$ and recovers PT in the thermodynamic limit. The EFE thus provides a deterministic bridge between geometry and criticality, revealing the geometric mechanisms that generate PTs.

Beyond its conceptual implications, the present framework has a universal scope and opens several directions for future investigation.
It applies naturally to finite systems such as proteins~\cite{koci2017subphase,schnabel2011microcanonical,bachmann2014thermodynamics,aierken2020comparison}, where the microcanonical ensemble provides the most direct thermodynamic description~\cite{dunkel2006phase}.
It also offers tools and insights for addressing long-range interacting models~\cite{campa2009statistical,dunkel2006phase,campa2014physics,dauxois2002dynamics,bouchet2010thermodynamics,defenu2023long,Giachetti2022Villain}, in which ensemble equivalence may break down and canonical methods or renormalization-group approaches may be limited.
In such cases, the geometric formulation provides a natural language to investigate the origin and structure of PTs.
When extended to include discrete configuration spaces, the framework can address spin systems where ensemble inequivalence also emerges~\cite{barre2001inequivalence,dauxois2000violation} and connects naturally with large-deviation formulations of non-equivalence~\cite{ellis2002nonequivalent,ellis2000large}.

\begin{acknowledgments}
The calculations presented in this paper were carried out using the HPC facilities of the University of Luxembourg~\cite{VBCG_HPCS14} {\small (see \href{http://hpc.uni.lu}{hpc.uni.lu})}.
\end{acknowledgments}

\clearpage
\onecolumngrid
\begin{center}\textbf{\Large Supplemental Material\\
Phase Transitions as Emergent Geometric Phenomena\\
 A Deterministic Entropy Evolution Law}\end{center}

\setcounter{secnumdepth}{2} 

\setcounter{section}{0}
\setcounter{equation}{0}
\setcounter{figure}{0}
\setcounter{table}{0}

\renewcommand{\thesection}{\Alph{section}}

\renewcommand{\thesubsection}{\thesection.\arabic{subsection}}

\setcounter{equation}{0}\setcounter{figure}{0}\setcounter{table}{0}
\renewcommand{\theequation}{S\arabic{equation}}
\renewcommand{\thefigure}{S\arabic{figure}}
\renewcommand{\thetable}{S\arabic{table}}

\section{Obstructions}
\label{sec:obstruction}

In this section, we show and discuss the obstruction of the symplectic form in naturally inducing a measure on the energy hypersurfaces. For further details about proofs and tools used in this section, we refer to Refs.~\cite{arnol2013mathematical,AbrahamMarsden,LeeSmooth,McDuffSalamon}.

\subsection{Degeneracy of the symplectic form on the energy hypersurfaces}\label{ssec:degeneracy}

We show that the restriction of the symplectic form $\omega$ to the tangent bundle of $\Sigma_E$ is degenerate, with the Hamiltonian vector field $\bm X_H$ in its kernel.\\

Let ${\bm x} \in \Sigma_E$ and consider the restriction $\omega|_{T_{\bm x} \Sigma_E}$, which we denote by $\omega_E$. For any vector $v \in T_x \Sigma_E$ (that is, $\bm v$ tangent to the level set $H = E$), we have, by definition:
\begin{equation}
dH(\bm v) = 0.
\end{equation}

By the defining equation of the Hamiltonian vector field, $\iota_{\bm X_H} \omega = dH$, we have for any vector $v$:
\begin{equation}
\omega(\bm X_H, \bm v) = dH(\bm v).
\end{equation}

In particular, for $v \in T_x \Sigma_E$:
\begin{equation}
\omega_E(\bm X_H,\bm v) = \omega(\bm X_H, \bm v) = dH(\bm v) = 0.
\end{equation}

Since this holds for all $\bm v \in T_{\bm x} \Sigma_E$, we have $\bm X_H \in \ker(\omega_E)$. The kernel is non-trivial; hence, $\omega_E$ is degenerate.

Moreover, $X_H(x) \in T_x \Sigma_E$ because
\begin{equation}
dH(\bm X_H) = \omega(\bm X_H, \bm X_H) = 0
\end{equation}
by the antisymmetry of $\omega$. Thus $X_H$ is both tangent to $\Sigma_E$ and in the kernel of $\omega$ restricted to $\Sigma_E$. 

\subsection{The Symplectic Complement and Coisotropic Submanifolds}
\label{ssec:coisotropy-complement}

Here, we show that the symplectic complement of $T_{\bm x} \Sigma_E$ is one-dimensional and is generated by $\bm X_H({\bm x})$. Furthermore, $(T_{\bm x} \Sigma_E)^{\perp_\omega} \subseteq T_{\bm x} \Sigma_E$, making $\Sigma_E$ a coisotropic submanifold.\\

Consider the symplectic complement, which is defined as
\begin{equation}
(T_{\bm x} \Sigma_E)^{\perp_\omega} = \{\bm w \in T_{\bm x} \Lambda : \omega(\bm w, \bm v) = 0 \; \forall \bm v \in T_{\bm x} \Sigma_E\}.
\end{equation}

We first show that $\bm X_H \in (T_{\bm x} \Sigma_E)^{\perp_\omega}$. For any $\bm v \in T_{\bm x} \Sigma_E$:
\begin{equation}
\omega(\bm X_H, \bm v) = dH(\bm v) = 0,
\end{equation}
as shown in Sec.~\ref{ssec:degeneracy}. Thus $\bm X_H \in (T_{\bm x} \Sigma_E)^{\perp_\omega}$.

To show that $\bm X_H$ generates the entire complement, we use a dimensional argument. For a subspace $W$ of a symplectic vector space $(V, \omega)$ of dimension $2n$, the symplectic complement satisfies
\begin{equation}
\dim(W) + \dim(W^{\perp_\omega}) = 2n.
\end{equation}

In our case, $\dim(T_{\bm x} \Sigma_E) = 2n - 1$ (since $\Sigma_E$ is a codimension-1 hypersurface), therefore:
\begin{equation}
\dim((T_{\bm x} \Sigma_E)^{\perp_\omega}) = 2n - (2n-1) = 1.
\end{equation}

Since $\bm X_H$ is non-zero (assuming $H$ is a regular function, i.e., $dH \neq 0$ on $\Sigma_E$) and belongs to $(T_{\bm x} \Sigma_E)^{\perp_\omega}$, and this complement has dimension $1$, we conclude:
\begin{equation}
(T_{\bm x} \Sigma_E)^{\perp_\omega} = \mathrm{span}\{X_H\}.
\end{equation}

Finally, we verify that $(T_x \Sigma_E)^{\perp_\omega} \subseteq T_x \Sigma_E$. We have already shown that $X_H \in T_x \Sigma_E$ (since $dH(X_H) = 0$). Thus, the symplectic complement is contained in the tangent space itself. This is precisely the defining property of a coisotropic submanifold \cite{McDuffSalamon}. 

\subsection{Non-uniqueness of Transverse Vectors}
\label{ssec:non-unique-transverse}
Finally, we show that the condition $dH(\bm\zeta) = 1$ does not uniquely determine a vector field $\bm\zeta$ transverse to $\Sigma_E$. The solution space has dimension $2n - 1$ at each point.\\

At each point $\bm x \in \Lambda$, we seek vectors $\bm\zeta \in T_{\bm x} \Lambda$ satisfying the linear equation
\begin{equation}
dH(\bm\zeta) = 1.
\end{equation}

This is a single linear equation in a $2n$-dimensional vector space. The solution set is an affine hyperplane:
\begin{equation}
S = \{\bm\zeta \in T_{\bm x} \Lambda : dH(\bm\zeta) = 1\}.
\end{equation}

To find the dimension, we first note that the solutions to the homogeneous equation $dH(\bm\zeta) = 0$ form the tangent space $T_{\bm x} \Sigma_E$, which has dimension $2n - 1$ \cite{LeeSmooth}.

The solution space $S$ to the inhomogeneous equation $dH(\bm\zeta) = 1$ is a translation of this vector space (an affine subspace); hence, it also has dimension $2n - 1$.

Explicitly, if $\bm\zeta_0$ is any particular solution (for example, $\bm\zeta_0 = \partial_{p_1}/(\partial H/\partial p_1)$ if $\partial H/\partial p_1 \neq 0$), then the general solution is:
\begin{equation}
\bm\zeta = \bm\zeta_0 + \bm v, \quad \text{where } \bm v \in T_{\bm x} \Sigma_E.
\end{equation}

Since $T_{\bm x} \Sigma_E$ has dimension $2n - 1$, we have a $(2n-1)$-parameter family of transverse vectors.

Different choices of $\bm\zeta$ yield different $(2n-1)$-forms $\iota_{\bm \zeta} d\mu_\Lambda$ when restricted to $\Sigma_E$. Without additional structure (such as a metric to define a canonical normal direction), there is no geometric principle within the symplectic framework to select one element of $S$ over another. 

\vspace{1em}
\noindent We remark that these three obstructions establish that symplectic geometry, while sufficient to identify the energy shell $\Sigma_E$ and govern the dynamics upon it, does not provide the geometric data necessary to measure the volume of $\Sigma_E$. The introduction of a metric tensor resolves all three obstructions simultaneously: it eliminates the dimensional mismatch by providing the gradient $\nabla H$ as a canonical transverse direction; it removes the non-uniqueness by defining orthogonality; and it repairs the degeneracy by inducing a non-degenerate metric on $\Sigma_E$ itself.

\section{Necessity of the energy clock}
\label{sec:energy-clock}
The gradient of $H$, while providing the transverse direction to $\Sigma_E$, does not provide a uniform parametrization of the energy increments across $\Sigma_E$. 

In order to see that consider a small displacement along $\nabla H$ under parametrization $\epsilon$: 
\[
\bm{x}(\epsilon)=\bm{x}_0+\nabla H(\bm{x}_0)\,d\epsilon\,.
\]
The target energy is 
\[
    E(\bm x(\epsilon))=H(\bm{x}_0)+\nabla H(\bm{x}_0)\cdot \nabla H(\bm{x}_0)\,d\epsilon.
\]
Therefore, the energy step is 
\[
    dE_{\bm x}=E(\bm{x}(\epsilon))-E_0=\|\nabla H(\bm{x}_0)\|^2\;d\epsilon
\]
Since $\|\nabla H\|$ varies across $\Sigma_E$, for two distinct points $\bm{x} \neq \bm{y}$ on the same $\Sigma_E$, we have $dE_{\bm{x}} \neq dE_{\bm{y}}$, mapping them to different target hypersurfaces. \\

To resolve this, it is sufficient to determine the vector field $\bm\zeta\in\text{span}\{\nabla H\}$ such that 
\[
    dH(\bm\zeta) = 1\implies \bm\zeta=\frac{\nabla_{\eta} H}{\|\nabla_{\eta} H\|_{\eta}^2}\,.
\]
This vector field generates the motion in energy that we call \textit{thermodynamic dynamics}, namely, the diffeomorphisms between energy hypersurfaces, 
\[
    \Phi^{\rm diff}_{\bm\xi}:\Sigma_E\to\Sigma_{E'},\qquad \frac{d\Phi^{\rm diff}}{d\epsilon}=\bm\zeta(\bm x(\epsilon))
\]
It thus provides the transverse direction that allows the decomposing 
\[
    T_{\bm x}\Lambda=\text{span}\{\bm\zeta\}\oplus T_{\bm x}\Sigma_E\,,
\]
and ensures that the energy motion maps all the points on $\Sigma_E$ into points of $\Sigma_{E'}$ simultaneously.

\section{Euclidean metric tensor in adapted frame}

The $\bm\zeta$-decomposition of $T_{\bm x}\Sigma_E$ naturally defines adapted coordinates $(E, y^\alpha)$ where $E = H(\bm x)$ parametrizes the energy and $\{y^\alpha\}_{\alpha=1}^{2n-1}$ on $\Sigma_E$.\\

Let $E:=H(\bm x)$ and let $y^\alpha$ ($\alpha=1,\dots,2n-1$) be local coordinates on the level set $\Sigma_E=\{H=E\}$. Choose the embedding $x=x(E,y)$ so that its coordinate vectors are
\[
    \partial_E x=\bm\zeta,\qquad \partial_\alpha x\in T\Sigma_E,\ \ \ dH(\partial_\alpha x)=0.
\]

In canonical coordinates $(q^i,p_i)$, the phase–space metric is
\[
    \eta=\delta_{ij}\,dq^i\!\otimes dq^j+\delta^{ij}\,dp_i\!\otimes dp_j.
\]
and in adapted coordinates $(E,y)$, its components become
\[
\eta_{AB}(E,y)\;=\;\eta\big(\partial_A x,\partial_B x\big),\qquad A,B\in\{E,\alpha\}.
\]

Then, using $dH(\partial_\alpha x)=0$ and $\bm\zeta=\nabla_\eta H/\|\nabla_\eta H\|_\eta^{2}$,
\begin{equation}\label{eqn:components-eta-adapted}
    \eta_{E\alpha}=\eta(\partial_E x,\partial_\alpha x)=\eta(\bm\zeta,\partial_\alpha x)
=\frac{dH(\partial_\alpha x)}{\|\nabla_\eta H\|_\eta^{2}}=0,\qquad
\eta_{EE}=\eta(\bm\zeta,\bm\zeta)=\frac{1}{\|\nabla_\eta H\|_\eta^{2}}.
\end{equation}
For the tangential components, we obtain the induced metric
\[
\sigma^{\eta}_{\alpha\beta}(E,y):=\eta(\partial_\alpha x,\partial_\beta x),
\]
where $\{\bm{e}_\alpha\}_{\alpha=1}^{2n-1}$ with $\bm{e}_\alpha:=\partial_\alpha x$ is the basis of $T_{\bm x}\Sigma_E$.

Therefore, the Euclidean metric in adapted coordinates takes the block–diagonal form
\begin{equation}\label{def:eta-adapted-coordinates}
        \eta\;=\;\frac{dE\otimes dE}{\|\nabla_\eta H\|_\eta^{2}}\;+\;\sigma^{\eta}_{\alpha\beta}(E,y)\,dy^\alpha\!\otimes dy^\beta\,,
\end{equation}
with $dE(\partial_E)=1$, $dE(\partial_\alpha)=0$, and $\sigma^\eta_E$ the metric induced on $\Sigma_E$.

Note that the microcanonical density of states naturally emerges as the induced measure on $\Sigma_E$.
Indeed, the Liouville measure in adapted coordinates is expressed as
\[
    d\mu^{\eta}_{\Lambda}=\frac{dEd\sigma_{E}}{\|\nabla_\eta H\|_{\eta}}=dp_1\ldots dp_n dq^1\ldots dq^n,\quad \implies    d\mu^{\eta}_E=d\mu^{\eta}_{\Lambda}\Big|_{\Sigma_E}=\frac{d\sigma^{\eta}_E}{\|\nabla_\eta H\|_{\eta}}
\]
which means
\begin{equation}
    \int_{\Sigma_E}d\mu_E=\int_{\Sigma_E}\frac{d\sigma_E}{\|\nabla_\eta H\|_{\eta}}=\int_{\Lambda}\delta(H-E)~dp_1\ldots dp_n dq^1\ldots dq^n=\Omega_{B}(E).
\end{equation}
where the coarea formula is used \cite{federer2014geometric}. We should remark on a subtle consequence due to the physical request behind the energy clock discussed in Sec.~\ref{sec:energy-clock}. 

From a pure mathematical viewpoint, given a generic oriented Riemannian manifold \((M,g)\) with volume form \(d\mu_X^g\), and smooth hypersurface \(\Sigma\subset M\), the area form is determined by the unit normal vector field $\bm{n}$:
(\(\|n\|_g=1\), \(g(n,T)=0\) for all \(T\in T\Sigma\)). This reads:
\[
d\sigma^g \;:=\; \iota_{n}\, d\mu_M^g\Big|_{\Sigma}.
\]
This is a coordinate-free definition (reversing \(n\) flips the orientation and the sign of the form),
and in adapted coordinates \((u^1,\dots,u^{m-1},u^m)\) with \(\Sigma=\{u^m=0\}\) it reduces to the usual
expression \(d\sigma^g=\sqrt{\det g_\top}\,du^1\wedge\cdots\wedge du^{m-1}\), where \(g_\top\) is the
metric induced on \(T\Sigma\). Then, the area of a Borel set \(A\subset\Sigma\) is \(\mathrm{area}_g(A)=\int_A d\sigma^g\).

\medskip
In our setting \((\Lambda,H,\omega,\eta)\), the energy shells are \(\Sigma_E=\{H=E\}\), and the geometric
area is
\[
\mathrm{area}_\eta(\Sigma_E)=\int_{\Sigma_E} d\sigma_E^{\eta},\qquad
d\sigma_E^{\eta}:=\sqrt{\det\sigma^{\eta}_E}\,dx^{1}\cdots dx^{2n-1}.
\]
However, thermodynamic quantities are described as function of energy, and therefore defined on $\Sigma_E$. Energy variation of a thermodynamic quantities must then be expressed in units of energy. Therefore, moving from one energy value to another of a quantity \(\delta E\) implies that we are following the energy motion identified by $\bm\zeta$. This vector indeed ``moves'' every point of \(\Sigma_E\) to \(\Sigma_{E+\delta E}\) as guaranteed by the condition $dH(\bm\zeta)=1$. 

Therefore, decomposing Liouville’s volume along this clock yields the canonical microcanonical surface measure
on \(\Sigma_E\) which is given by
\[
d\mu_E^{\eta}:=\iota_{\zeta}\, d\mu_{\Lambda}\Big|_{\Sigma_E}
=\frac{\iota_{n}d\mu_{\Lambda}}{\|\nabla_{\eta}H\|_{\eta}}
=\frac{d\sigma_E^{\eta}}{\|\nabla_{\eta}H\|_{\eta}}.
\]

\section{Mapping from Euclidean metric tensor to unit-norm gauge tensor}

Let us consider the Euclidean metric tensor in adapted coordinates in Eq.~\eqref{def:eta-adapted-coordinates}. We define the transformation \cite{gori_configurational}:
\begin{equation}\label{eq:rescaling-coords}
    dx^{0} = \chi\, dE, \quad
    dx^{i} = \chi^{-\frac{1}{n-1}}\, dy^{i}, \qquad\chi := \frac{1}{\|\nabla H\|_\eta}
\end{equation}
where $i\in[1,2n-1]$. In these new coordinates $(x^0,x^1,\ldots,x^{2n-1})$, the metric tensor is (bi-)conformally related to $\eta$:
\begin{equation}\label{eq:biconformal-components}
    g_{00} = \chi^{-2}\,\eta_{EE}, \qquad
    (\sigma^g_E)_{ij} = \chi^{\frac{2}{n-1}}\, (\sigma^\eta_E)_{ij},
\end{equation}
yielding the equivalent form
\begin{equation}\label{eq:unit-norm-metric}
    g = dx^{0}\otimes dx^{0} + (\sigma^g_E)_{ij}\, dx^{i}\otimes dx^{j}.
\end{equation}

This defines a new metric tensor $g$ that we call the \textit{unit-norm gauge}. This metric tensor has peculiar properties. 

\subsection{The vector field generating the energy flow in the unit-norm gauge}

First, in this metric, the vector field generating the flow of the diffeomorphism $\bm\xi$ must satisfy $dx^0(\bm\xi)=1$, and therefore $\bm\xi\equiv\partial_{x^0}$. From Eq.~\eqref{eq:rescaling-coords}, we have 
\begin{equation}
    \begin{split}\label{def:clock-xi}
            dx^0(\bm\xi)=\chi dE(\bm\xi)=1,
    \end{split}
\end{equation}
on the other side, $dE(\bm\zeta)=1$ therefore
\[
    dE(\bm\zeta)=dx^0(\bm\xi)=1\quad\implies \bm\xi=\chi^{-1}\bm\zeta.
\]
Finally, $\bm\xi$ is such that $\|\bm\xi\|_g=1$. This is easy to check
\begin{equation}
    \begin{split}\label{def:unit-norm-xi}
        g(\bm\xi,\bm\xi)=g(\partial_{x^0},\partial_{x^0})=g_{00}=\chi^{-2}\eta_{EE}=1,
    \end{split}
\end{equation}
as follows from Eq.~\eqref{eqn:components-eta-adapted}.

Finally, we observe that $\bm\xi=\nabla_g H$. Indeed:
\[
    \nabla_g H=g^{00}\frac{\partial H}{\partial x^0}\partial_{x^0}+g^{ij}\frac{\partial H}{\partial x^i}\partial_{x^j}
\]
However, independently by the metric tensor ($g$ remains compatible with $\omega$):
\[
    g(\nabla_g H,\bm{X}_H)=dH(\bm{X}_H)=0
\]
therefore, any components of $\nabla_g H$ on $T_{\bm x}\Sigma_E$ must vanish: $\partial H/\partial x^i=0$ which leads to
\[
    \nabla_gH=g^{00}\frac{\partial H}{\partial x^0}\partial_{x^0}
\]
Finally, since $g^{00}=g_{00}=1$, we notice that 
\[
    \frac{\partial H}{\partial x^0}=dH(\partial_{x^0})=1\;\;\implies \nabla_g H\equiv\bm\xi\equiv\partial_{x^0}.
\]

Finally, we notice that in the unit-norm gauge the induced measure on $\Sigma_E$ coincides with the area form
\(d\mu_E^{g}=d\sigma_E^{g}\), being $\|\bm\xi\|_g=1$. Then, the Boltzmann density of states coincides with the geometric area:
\[
\mathrm{area}_g(\Sigma_E)=\int_{\Sigma_E} d\sigma_E^{g}=\int_{\Sigma_E} \frac{d\sigma^\eta_E}{\|\nabla_\eta H\|_\eta^2}=\int_{\Lambda}\delta(H-E)\,d\mu_{\Lambda}=\Omega_B(E).
\]

\section{Calculation of second-order GCF}
\label{app:calculation-2-GCF}

In this section, we compute the explicit form of $\Upsilon^{(2)}_g$ in terms of gradient, Hessian and Laplacian associated to the metric tensor $\eta$. The expression that we will obtain is the one used in numerical simulations, namely, evaluated during the dynamics. We consider
\begin{equation}
    \mathrm{Tr}[W_{\bm \xi}]
    =\frac{\Delta H}{\|\nabla H\|^{2}}
     -2\,\frac{\langle\nabla H,(\text{Hess}\,H)\,\nabla H\rangle}{\|\nabla H\|^{4}},
    \qquad
    \bm{\xi}=\frac{\nabla H}{\|\nabla H\|^{2}},
\end{equation}
we introduce the shorthand notation:
\[
    u:=\nabla H,\qquad 
    G:=\|\nabla H\|^{2},\qquad
    A:=\Delta H,\qquad
    B:=\langle u,(\text{Hess}\,H)u\rangle.
\]
Then,
\begin{equation}
    \mathrm{Tr}[W_{\bm \xi}]=\frac{A}{G}-\frac{2B}{G^{2}}.
\end{equation}
Since $\partial_{E}=\mathscr{L}_{\bm{\xi}}=(\bm{\xi}\!\cdot\!\nabla)=(u/G)\!\cdot\!\nabla$, we can compute $\partial_{E}\mathrm{Tr}[W_{\bm \xi}]$, i.e.:
\begin{align}
    \partial_{E}\mathrm{Tr}[W_{\bm \xi}]
    &=\frac{\langle u,\nabla A\rangle}{G^{2}}
      -\frac{2A\langle u,\nabla G\rangle}{G^{3}}
      -\frac{2\langle u,\nabla B\rangle}{G^{3}}
      +\frac{4B\langle u,\nabla G\rangle}{G^{4}}.
\end{align}
Using the auxiliary relations
\[
\langle u,\nabla G\rangle = 2B, \qquad 
\langle u,\nabla B\rangle = 2\,\langle u,(\text{Hess}\,H)^{2}u\rangle
+\nabla^{3}H(u,u,u),
\]
we obtain
\begin{equation}
    \partial_{E}\mathrm{Tr}[W_{\bm \xi}]=\frac{\langle u,\nabla A\rangle}{G^{2}}
    -\frac{2AB}{G^{3}}-\frac{4\,\langle u,(\text{Hess}\,H)^{2}u\rangle+2\,\nabla^{3}H(u,u,u)}{G^{3}}+\frac{8B^{2}}{G^{4}}.
\end{equation}
Furthermore,
\[
    \big(\mathrm{Tr}[W_{\bm \xi}]\big)^{2}
   =\frac{A^{2}}{G^{2}}
    -\frac{4AB}{G^{3}}
    +\frac{4B^{2}}{G^{4}}\,.
\]
Therefore:
\[
    \partial_{E}\mathrm{Tr}[W_{\bm \xi}]+\big(\mathrm{Tr}[W_{\bm \xi}]\big)^{2}=
   \frac{A^{2}+\langle u,\nabla A\rangle}{G^{2}}
   -\frac{6AB+4\,\langle u,(\text{Hess}\,H)^{2}u\rangle
      +2\,\nabla^{3}H(u,u,u)}{G^{3}}
   +\frac{12B^{2}}{G^{4}}
\]
Collecting terms gives the final result:
\begin{equation}
\begin{aligned}
\partial_{E}^{2}{\rm vol}^{g}(E)
   &=\int_{\Sigma_E}\!\!
   \Bigg[
   \frac{A^{2}+\langle u,\nabla A\rangle}{G^{2}}
   -\frac{6AB+4\,\langle u,(\text{Hess}\,H)^{2}u\rangle
      +2\,\nabla^{3}H(u,u,u)}{G^{3}}
   +\frac{12B^{2}}{G^{4}}
   \Bigg]d\mu_g\\[4pt]
   &=\int_{\Sigma_E}\!\!
   \Bigg[
   \frac{(\Delta H)^{2}+\langle\nabla H,\nabla(\Delta H)\rangle}{\|\nabla H\|^{4}}
   -\frac{
      6\,(\Delta H)\langle\nabla H,(\text{Hess}\,H)\nabla H\rangle}{\|\nabla H\|^{6}}-\frac{
      4\,\langle\nabla H,(\text{Hess}\,H)^{2}\nabla H\rangle
      }{\|\nabla H\|^{6}}\\
      &\hspace{5cm}-\frac{2\,\nabla^{3}H(\nabla H,\nabla H,\nabla H)
      }{\|\nabla H\|^{6}}
   +\frac{
      12\,\langle\nabla H,(\text{Hess}\,H)\nabla H\rangle^{2}
      }{\|\nabla H\|^{8}}
   \Bigg]d\mu_g.
\end{aligned}
\end{equation}
Finally, the second-order GCF reads
\begin{equation}
    \begin{split}
\Upsilon^{(2)}_g(E)=\frac{\partial_{E}^{2}{\rm vol}^{g}(E)}{\text{area}^g(E)}
   &=\int_{\Sigma_E}\!\!
   \Bigg[
   \frac{(\Delta H)^{2}+\langle\nabla H,\nabla(\Delta H)\rangle}{\|\nabla H\|^{4}}
   -\frac{
      6\,(\Delta H)\langle\nabla H,(\text{Hess}\,H)\nabla H\rangle}{\|\nabla H\|^{6}}-\frac{
      4\,\langle\nabla H,(\text{Hess}\,H)^{2}\nabla H\rangle
      }{\|\nabla H\|^{6}}\\
      &\hspace{5cm}-\frac{2\,\nabla^{3}H(\nabla H,\nabla H,\nabla H)
      }{\|\nabla H\|^{6}}
   +\frac{
      12\,\langle\nabla H,(\text{Hess}\,H)\nabla H\rangle^{2}
      }{\|\nabla H\|^{8}}
   \Bigg]d\rho_g.
    \end{split}
\end{equation}
with $d\rho_g=d\mu^g_E/\text{area}^g(E)$. Considering the microcanonical average 
\[
    \langle O\rangle_E=\int O(\pi,\phi)\delta(H(\pi,\phi)-E)~D\pi\,D\phi
\]
then this object is estimated by computing
\begin{equation}\label{def:estimation-GCF}
    \begin{split}
\Upsilon^{(2)}_g(E)=\left\langle  \frac{(\Delta H)^{2}+\langle\nabla H,\nabla(\Delta H)\rangle}{\|\nabla H\|^{4}}\right\rangle_E&-\left\langle\frac{
      6\,(\Delta H)\langle\nabla H,(\text{Hess}\,H)\nabla H\rangle}{\|\nabla H\|^{6}}\right\rangle_E-\left\langle\frac{
      4\,\langle\nabla H,(\text{Hess}\,H)^{2}\nabla H\rangle
      }{\|\nabla H\|^{6}}\right\rangle_E\\
      &-\left\langle\frac{2\,\nabla^{3}H(\nabla H,\nabla H,\nabla H)
      }{\|\nabla H\|^{6}}\right\rangle_E
   +\left\langle\frac{
      12\,\langle\nabla H,(\text{Hess}\,H)\nabla H\rangle^{2}
      }{\|\nabla H\|^{8}}\right\rangle_E
    \end{split}
\end{equation}

\section{Expansion of the Weingarten operator}\label{}

\subsection{Angular statistics and small-$M$ expansions}\label{app:angular-averages}

Let $\Theta_i=\theta_i-\phi$ and $M:=\big\|\frac{1}{N}\sum_i(\cos\theta_i,\sin\theta_i)\big\|\ll 1$. On the disordered branch we model the angular fluctuations by the von Mises density
\[
f(\Delta)=\frac{e^{h\cos\Delta}}{2\pi I_0(h)},\qquad h\ll 1,
\]
whose moments satisfy (see standard Refs.~\cite{MardiaJupp,Fisher93,Jammalamadaka2001})
\[
    \langle \cos(k\Delta)\rangle = I_k(h)/I_0(h)
\]
Another standard identity is $\langle \sin(k\Delta)\rangle=0$ by symmetry, whence
\[
\langle \cos\Delta\rangle=\frac{I_1}{I_0}= \frac{h}{2}-\frac{h^3}{16}+O(h^5)=:M
\quad\Rightarrow\quad h=2M+O(M^3).
\]
Small–$h$ expansions for the modified Bessel functions \(I_k\) follow from classical handbooks \cite{AbramowitzStegun}.
Using $\cos^2\Delta=\tfrac12(1+\cos2\Delta)$ and the small-$h$ expansions of $I_k$,
\[
\langle \cos^2\Delta\rangle=\frac12+\frac{1}{2}\frac{I_2}{I_0}
=\frac12+\frac{h^2}{16}+O(h^4)
=\frac12+\frac{M^2}{4}+O(M^4),
\]
hence
\begin{equation}
\langle \sin^2\Delta\rangle=\frac12-\frac{M^2}{4}+O(M^4).
\label{eq:sin2-avg}
\end{equation}
Moreover, using $\cos^3\Delta=\tfrac14(3\cos\Delta+\cos3\Delta)$,
\[
\langle \cos^3\Delta\rangle=\frac{3}{4}M+\frac{1}{4}\frac{I_3}{I_0}
=\frac{3}{4}M+\frac{h^3}{192}+O(h^5)
=\frac{3}{4}M+\frac{M^3}{24}+O(M^5),
\]
so that
\begin{equation}
\langle \sin^2\Delta\,\cos\Delta\rangle
=\langle \cos\Delta-\cos^3\Delta\rangle
=\frac{1}{4}M-\frac{1}{24}M^3+O(M^5).
\label{eq:sin2cos-avg}
\end{equation}

By the law of large numbers, sums over $i=1,\dots,N$ are replaced by $N$ times the expectations up to $O(\sqrt{N})$ fluctuations. Therefore,
\begin{equation}
    S_2:=\sum_{i=1}^N \sin^2\Theta_i= N\,\langle \sin^2\Delta\rangle= N\Big(\frac12-\frac14 M^2\Big)+O(NM^4)+O(\sqrt{N}),
\label{eq:S2-exp}
\end{equation}
\begin{equation}
    S_{21}:=\sum_{i=1}^N \sin^2\Theta_i\,\cos\Theta_i= N\,\langle \sin^2\Delta\,\cos\Delta\rangle= \frac{MN}{4}+O(NM^3).
\label{eq:S21-exp}
\end{equation}

For the scaling of $S_{\mathrm{int}}$
we use $\cos(\Theta_i-\Theta_j)=\cos\Theta_i\cos\Theta_j+\sin\Theta_i\sin\Theta_j$, and we get
\begin{align}
S_{\mathrm{int}}
&=\sum_{i\neq j}\sin\Theta_i\sin\Theta_j\cos(\Theta_i-\Theta_j)\nonumber\\
&=\sum_{i\neq j}\big(\sin\Theta_i\cos\Theta_i\big)\big(\sin\Theta_j\cos\Theta_j\big)
 +\sum_{i\neq j}\sin^2\Theta_i\,\sin^2\Theta_j \nonumber\\
&=\Big(\sum_i \sin\Theta_i\cos\Theta_i\Big)^2-\sum_i \sin^2\Theta_i\cos^2\Theta_i
 +\Big(\sum_i \sin^2\Theta_i\Big)^2-\sum_i \sin^4\Theta_i. \label{eq:Sint-decomp}
\end{align}
By symmetry $\sum_i \sin\Theta_i\cos\Theta_i=O(\sqrt{N})$, hence its square is $O(N)$ and subleading versus $N^2$ terms.
Set $A:=\sum_i\sin^2\Theta_i$; from \eqref{eq:S2-exp}, \[
    A=\frac{N}{2}-\frac{N}{4}M^2+O(NM^4)
\]
Using 
\[
    \sin^4\Delta=\frac{3-4\cos 2\Delta+\cos 4\Delta}{8}
\]
and $I_4/I_0=O(h^4)$, we get
\[
\langle \sin^4\Delta\rangle=\frac{3}{8}-\frac{1}{2}\frac{I_2}{I_0}+\frac{1}{8}\frac{I_4}{I_0}
=\frac{3}{8}-\frac{h^2}{16}+O(h^4)
=\frac{3}{8}-\frac{M^2}{4}+O(M^4),
\]
so that $\sum_i\sin^4\Theta_i=N\big(\tfrac{3}{8}-\tfrac{M^2}{4}\big)+O(NM^4)+O(\sqrt{N})$.
Plugging into \eqref{eq:Sint-decomp} and retaining $O(N^2)$ terms,
\begin{equation}
S_{\mathrm{int}}
=\Big(\tfrac{N}{2}-\tfrac{N}{4}M^2\Big)^2 - \sum_i \sin^4\Theta_i + O(N)
=\frac{N^2}{4}+O(N^2 M^2) - \frac{3N}{8}+O(N),
\label{eq:Sint-scaling}
\end{equation}
which yields, to leading order in $N$,
\begin{equation}
S_{\mathrm{int}}\simeq \frac{N^2}{4}\qquad (N\to\infty,\ M\to 0).
\end{equation}

\subsection{Denominator expansions and block sums}\label{app:denominators}

Recall $K:=\sum_i p_i^2$ and
\[
G:=\|\nabla H\|^2 = K + J^2 M^2 S_2.
\]
Write $c:=K/N$ (finite in the thermodynamic limit). Using \eqref{eq:S2-exp},
\[
G = N\Big(c+\frac{J^2}{2}M^2\Big)+O(NM^4).
\]
Therefore,
\begin{align}
\frac{1}{G}
&=\frac{1}{Nc}\Big(1+\frac{J^2}{2c}M^2\Big)^{-1}
=\frac{1}{Nc}\Big(1-\frac{J^2}{2c}M^2+O(M^4)\Big), \label{eq:oneoverG}\\
\frac{1}{G^2}
&=\frac{1}{N^2c^2}\Big(1+\frac{J^2}{2c}M^2\Big)^{-2}
=\frac{1}{N^2c^2}\Big(1-\frac{J^2}{c}M^2+O(M^4)\Big).\label{eq:oneoverG2}
\end{align}

Inserting \eqref{eq:S2-exp}, \eqref{eq:S21-exp}, \eqref{eq:Sint-scaling}, \eqref{eq:oneoverG}–\eqref{eq:oneoverG2} into the blocks
\[
\sum_i\kappa_{p_i}=\frac{N}{G}-\frac{2K}{G^2},\qquad
\sum_i\kappa^{\mathrm{d}}_{\theta_i}=\frac{J N M^2}{G}-\frac{2J^3 M^3 S_{21}}{G^2},\qquad
\sum_{i\neq j}\kappa^{\mathrm{int}}_{ij}=\frac{2J^3 M^2}{N\,G^2}S_{\mathrm{int}},
\]
and keeping only $O(N^0)$ terms (the ones that survive as $N\to\infty$), reproduces the expressions for $\bm P$ and $\bm I$ of the main text.

\medskip

\noindent\textbf{Summary (small-$M$, thermodynamic limit).}
\[
S_2\simeq N\Big(\frac12-\frac14 M^2\Big),\qquad
S_{21}\simeq \frac{MN}{4},\qquad
S_{\mathrm{int}}\simeq \frac{N^2}{4},
\]
\[
\frac{1}{G}\simeq \frac{1}{Nc}\Big(1-\frac{J^2}{2c}M^2\Big),\qquad
\frac{1}{G^2}\simeq \frac{1}{N^2c^2}\Big(1-\frac{J^2}{c}M^2\Big),
\]

\section{Numerical methods and integration schemes}\label{sec:numerical-sampling}

Consider $\phi$ as a generalized coordinates: this can be the real field configuration in the 2D $\phi^4$ model or the angle variable $\theta$ in the XY model.\\

\subsection{Configurational microcanonical Monte Carlo algorithm}
\label{app:micro_montecarlo}
The microcanonical ensemble has been reproduced by adopting the method proposed in Refs.~\cite{ray1996microcanonical,ray1991microcanonical} and consisting of a microcanonical Monte Carlo (MICROMC) algorithm in the configuration space. We have chosen a method based on a random proposal to enforce ergodicity. 

Note that the numerical procedure that we adopted and reported below is system-independent.

\subsection{Microcanonical distribution function}
The microcanonical algorithm is based on the identification of a suitable density probability function that is used to produce a reliable sampling. To find such a probability function, we start with the microcanonical partition function:
\[
\Omega(E) = \int \delta(H[\pi, \phi] - E) D\pi\,D\phi\,,
\]
where
\[
    D\pi=\prod_{\bm{n}\in\mathbb{L}}d\pi_{\bm{n}},\qquad D\phi=\prod_{\bm{n}\in\mathbb{L}}d\phi_{\bm{n}}\,,
\]
and where the Hamiltonian is
\[
    H(\pi, \phi) = \sum_{\bm{n}\in\mathbb{L}} \frac{\pi_{\bm{n}}^2}{2}  + V(\phi)\,,
\]
for any potential function. Finally, $\delta$ represents the Dirac delta function. From now on, we introduce
$L=N^2$ to denote the total number of degrees of freedom. We focus on the momentum integral that separates as

\begin{equation}\label{def:integral_momentum_micro}
    I_p = \int \delta \left(\sum_{\bm{n}\in\mathbb{L}} \frac{\pi_{\bm{n}}^2}{2} + V(\phi) - E\right)D\pi\,.
\end{equation}
Thus, we notice that the kinetic energy variable,
\[
    K = \sum_{\bm{n}\in\mathbb{L}} \frac{p_{\bm{n}}^2}{2}\;,
\]
can be interpreted as the equation for a $L$-dimensional sphere of radius $r^2=2K$. Then, exploiting the spherical coordinates in $L$ dimensions, we have
\[
    \prod_{\bm{n}\in\mathbb{L}}d\pi_{\bm{n}} = \rho^{L-1}\,d\rho\,d\Omega_{L-1}\,,
\]
where \( d\Omega_{L-1} \) is the differential solid angle element on the unit sphere \( S^{L-1} \), expressed as:
\[
    d\Omega_{L-1} = \prod_{i=1}^{L-1} d\theta_i \sin^{L-1-i} \theta_i,\qquad
    \Omega_{L}=\int d\Omega_{L-1}\;.
\]
Then, we define 
\[
    \rho=\sqrt{2K},\quad d\rho=\frac{dK}{\sqrt{2K}}\implies \rho^{L-1}d\rho=(2K)^{L/2-1}\,dK\,.
\]
Finally, integral \eqref{def:integral_momentum_micro} rewrites
\begin{equation}\label{}
\begin{split}
    I_p &= \Omega_{L}\int \delta \left[K-(E- V(\phi))\right](2K)^{L/2-1}\;dK\\
    &= 2^{L-1}\Omega_{L}(E- V(\phi))^{L/2-1}\Theta(E - V(\phi))\,.
\end{split}
\end{equation}
Reintroducing above the integral over field configurations, we recover the \emph{configurational microcanonical partition function} that reads
\begin{equation}\label{def:configurational_entropy}
    \Omega(E) = \int  (E - V(\phi))^{L/2 - 1} \Theta(E - V(\phi))\;D\phi.
\end{equation}
Notice that we have dropped irrelevant constants that play no role in the calculation of expectation values.\\

The numerical algorithm then exploits the probability distribution function arising from Eq.~\eqref{def:configurational_entropy}
$$
    W_{E}[\phi]= \left(E-V(\phi)\right)^{L/2-1}\Theta[E-V(\phi)]\,.
$$
The sampling is carried out by randomly picking a lattice site, $\bm{n}$, and proposing a new random configuration $\phi^{old}_{\bm{n}}\mapsto \phi^{new}_{\bm{n}}=\phi^{old}_{\bm{n}} + \eta\Delta\phi$ where $\eta$ is a random number uniformly sampled from the interval $[-1,1]$ and $\Delta\phi$ is adjusted to achieve a target acceptance rate of $50\%-60\%$. The new configuration is accepted according to the Monte Carlo acceptance probability
$$
    W(\phi^{old}\to\phi^{new}) = \min\left(1,\frac{W_{E}[\phi^{new}]}{W_{E}[\phi^{old}]}\right)\,.
$$
Notice that an efficient way to evaluate this acceptance probability is to rewrite the acceptance ratio as follows:
$$
    \frac{W_{E}[\phi^{new}]}{W_{E}[\phi^{old}]} = \exp\left[\bigg(\frac{L}{2}-1\bigg)\log\left(\frac{E-V(\phi^{new})}{E-V(\phi^{old})}\right)\right]\,.
$$
once we have checked that $E-V(\phi^{new})>0$.

\subsection{Initial configuration}

Initial conditions are randomly proposed in order to start with a high-energy configuration, $E_{rand}$, larger than the desired input energy $E_{inp}$ and such that $E_{rand}-V(\phi_{ini})>0$. Then, we perform $10^{4}$ steps of equilibration with the MICROMC algorithm. At this stage, we search for the correct value for $\Delta\phi$. To do that, we start with a small value for $\Delta\phi$, say $0.001$ and run a few MICROMC steps using the desired input energy $E_{inp}$ and computing the acceptance rate, $N_{\text{acc}}$. If $N_{\text{acc}}\notin[0.5,0.6]$, then we repeat the procedure by replacing $\Delta\phi$ with $\Delta\phi+0.001$. It should be stressed that other strategies, such as the molecular dynamics algorithm, have also been used to select the initial conditions. Comparison of the thermodynamic observables obtained with both methods did not yield appreciable differences. Finally, once the suitable tune parameter's value has been obtained, the configuration is equilibrated for $10^4$ steps, and the trajectory is evolved with the MICROMC method and used for computing averages. For each energy value, we have produced $N_{trj}=32$ realizations for each system size. Each thermodynamic observable has been evaluated through $N_{\text{avg}}=10^{6}$ measurements, performed in every $N_{\text{step}}=100$ MICROMC step. The microcanonical average of a given observable, $f$, is computed by
\begin{equation}\label{def:time_average}
    \langle f\rangle_{\varepsilon}=\frac{1}{ N_{\text{trj}}\cdot N_{\text{avg}}}\sum_{i=1}^{N_{trj}}\sum_{\alpha=1}^{N_{\text{avg}}}f(\phi^{(i)}_{\alpha})\,,
\end{equation}
where $f$ is the observable evaluated on the configuration $\phi^{(i)}_{\alpha}$ at the $\alpha$-th MC step for the $i$-th realization.

\subsection{Numerical calculation of microcanonical entropy's derivatives}
\label{app:PHT_method}
The MIPA method requires the estimation of higher-order derivatives of the microcanonical entropy from a microcanonical sampling of the phase space. To this purpose, we adopt the Pearson-Halicioglu-Tiller (PHT) method \cite{pearson1985laplace} which allows a direct calculation of the first- and second-order derivatives of the microcanonical entropy as follows. Any energy derivative of the entropy function can be rewritten in terms of averages of power of the kinetic energy $k=(E-V(\phi)/L$. In this framework, the first- and second-order derivatives of the microcanonical entropy can be written as 
\begin{align}
     \partial_{\varepsilon}S(\varepsilon)&= \left(\frac{1}{2}-\frac{1}{L}\right)\langle k^{-1}\rangle_{\varepsilon}\,, \label{eqn:derS-I}\\
    \partial^{2}_{\varepsilon}S(\varepsilon)
    &=L\bigg[\bigg(\dfrac{1}{2}-\dfrac{1}{L}\bigg) \bigg(\dfrac{1}{2}-\dfrac{2}{L}\bigg)\langle k^{-2} \rangle_{\varepsilon}
    - \bigg(\dfrac{1}{2}-\dfrac{1}{L}\bigg)^2 \langle k^{-1}\rangle^2_{\varepsilon} \bigg]\,, \label{eqn:derS-II}
\end{align}
where the microcanonical averages have been estimated by Eq.~\eqref{def:time_average} over the realizations.

\section{Numerical solution of the Entropy Flow Equation (EFE)}\label{sec:efe-solution-procedure}

\paragraph{Equation and unknowns.}
We solve the EFE
\begin{equation}\label{eq:EFE}
    S_g''(E)+\big(S_g'(E)\big)^2 \;=\; \Upsilon^{(2)}_g(E),
\end{equation}
on an interval \(E\in[E_{\min},E_{\max}]\).
Set
\[
    g(E):=S_g'(E),\qquad S_g'(E)=g(E),\qquad S_g''(E)=g'(E),
\]
so that \eqref{eq:EFE} reduces to the Riccati ODE
\begin{equation}\label{eq:Riccati}
    g'(E)+g(E)^2 \;=\; \Upsilon^{(2)}_g(E).
\end{equation}
Once \(g\) is known, the entropy follows by a single quadrature,
\begin{equation}\label{eq:S-from-g}
    S_g(E)=S_g(E_0)+\int_{E_0}^{E} g(\tilde E)\,d\tilde E.
\end{equation}

\paragraph{Input data: geometric source.}
The source \(\Upsilon^{(2)}_g\) is computed numerically at a discrete set of energies
\(\{(E_i,\Upsilon_i)\}_{i=1}^M\), with \(E_{\min}\le E_1<\dots<E_M\le E_{\max}\).
To obtain a smooth right–hand side in \eqref{eq:Riccati}, we construct a high–order
interpolant
\begin{equation}\label{eq:interp}
    \widehat\Upsilon(E)\;:=\;\mathcal{I}_{\!p}\big[\{(E_i,\Upsilon_i)\}\big](E),
\end{equation}
using a polynomial spline of order \(p=9\) (Mathematica’s \texttt{Interpolation} with \texttt{InterpolationOrder\(\to\)9}).
This yields a \(C^{p-1}\) function \(\widehat\Upsilon\) on \([E_{\min},E_{\max}]\) that we use in place of \(\Upsilon^{(2)}_g\) in \eqref{eq:Riccati}.

\paragraph{Initial data.}
We pick a left endpoint \(E_0\in[E_{\min},E_{\max})\) and prescribe:
\begin{equation}\label{eq:ICs}
    g(E_0)=\beta_0,\qquad S_g(E_0)=0,
\end{equation}
where \(\beta_0\) is the microcanonical inverse temperature at \(E_0\) obtained from numerical simulation through Eq.~\eqref{eqn:derS-I} and $S_g(E_0)$ sets the additive entropy, since it cancels in derivatives, it can be fixed to zero.

\paragraph{Time–integration (energy stepping).}
We integrate the first–order system
\begin{equation}\label{eq:first-order-system}
\begin{cases}
S'(E)=g(E),\\[2pt]
g'(E)=\widehat\Upsilon(E)-g(E)^2,
\end{cases}
\qquad E\in[E_0,E_{\max}],
\end{equation}
with an explicit high–order Runge–Kutta method. In practice we used Mathematica’s
\texttt{NDSolve} with
\[
\texttt{Method}\;=\;\{\texttt{"TimeIntegration"}\to\{\texttt{"ExplicitRungeKutta"},\ \texttt{"DifferenceOrder"}\to 9\}\},
\]
and uniform step \(\Delta E\) on \([E_0,E_{\max}]\).
The numerical solution returns \(S(E)\), \(g(E)\), and, where needed, \(g'(E)\).

\paragraph{Discretization and sampling.}
For visualization and post–processing we evaluate \((g,g')\) on the uniform grid
\[
E_j \;=\; E_0 + j\,\Delta E,\qquad j=0,1,\dots,J,\quad E_J\le E_{\max}.
\]
This produces the arrays \(\{(E_j,g(E_j))\}\) and \(\{(E_j,g'(E_j))\}\) used in the figures.

\paragraph{Stability and cross–checks.}
The Riccati form \eqref{eq:Riccati} can be validated against the equivalent second–order
\emph{linear} problem via the substitution \(g=\psi'/\psi\), namely
\begin{equation}\label{eq:linear-check}
\psi''(E)\;=\;\widehat\Upsilon(E)\,\psi(E),\qquad
g(E)=\frac{\psi'(E)}{\psi(E)},\qquad
S_g(E)=S_0+\ln\frac{\psi(E)}{\psi(E_0)}.
\end{equation}
We solved \eqref{eq:linear-check} with the same RK scheme and verified that the reconstructed \(g\) matches the direct Riccati integration within the stated tolerances. Convergence was monitored by halving \(\Delta E\) and by lowering the interpolant order \(p\) (from 9 to 7) to ensure insensitivity to the smoothing.

\paragraph{Reproducibility notes.}
(i) \(\widehat\Upsilon(E)\) is evaluated only within the range \(\{E_i\}\) used in numerical simulations (no extrapolation).
(ii) All ODE solves use a fixed step and order (RK9); decreasing \(\Delta E\) by a factor of 2 changes \(g\) and \(g'\) by less than the line width.
(iii) The additive constant \(S_0\) is fixed by matching \(S_g\) to the reference at \(E_0\) (or by setting \(S_0=0\)); derivatives are independent of \(S_0\).

\paragraph{Output.}
The procedure returns smooth functions \(g(E)=S_g'(E)\) and \(g'(E)=S_g''(E)\) on \([E_0,E_{\max}]\), together with the entropy \(S_g(E)\) from \eqref{eq:S-from-g}. These are the curves reported in the figures and used to compare with microcanonical observables.

\section{Investigation of phase transitions}

The Hamiltonian for the 2D $\phi^4$ model with nearest neighbor interactions is 
\begin{equation}
\begin{split}\label{def:hamiltonian_phi4}
    H:=\sum_{\bm{i}}\Bigg[\frac{\pi^{2}_{\bm{i}}}{2}+\frac{\lambda}{4!}\phi^{4}_{\bm{i}}&-\frac{\mu^{2}}{2}\phi^{2}_{\bm{i}}+\frac{J}{4}\sum_{\bm{k}\in N(\bm{i})}(\phi_{\bm{i}}-\phi_{\bm{i}})^{2}\Bigg]
\end{split}
\end{equation}
and we choose values 
\[
    \lambda=3/5,\qquad \mu=2,\qquad J=1.
\]
The label $\bm{i}:=(i_1,i_2)$ is the two-dimensional array of integer numbers used for labeling the sites; finally, we denote with $N(\bm{i})$ the set of the nearest neighbors of the $\bm{i}$-th site.

The Hamiltonian of the 1D XY mean-field is given by
\begin{equation}
    H(\theta,p)=\sum_{i=1}^N \frac{p_i^2}{2}+\frac{J}{2N}\sum_{i,j}\bigl[1-\cos(\theta_i-\theta_j)\bigr],
\end{equation}
where $J=1$.

For the $\phi^4$-model we chose the the energy range $[9,16]$ with energy resolution $\Delta E=0.1$ and system sizes $N=64^4,\,128^2,\,256^2$.
For XY model, the energy range is $[0.5,1]$ with energy resolution $\Delta E=0.01$ and studied three different system sizes $N=8100,\,14400,40000$. 

\begin{figure*}
    \centering
    \includegraphics[width=0.9\linewidth]{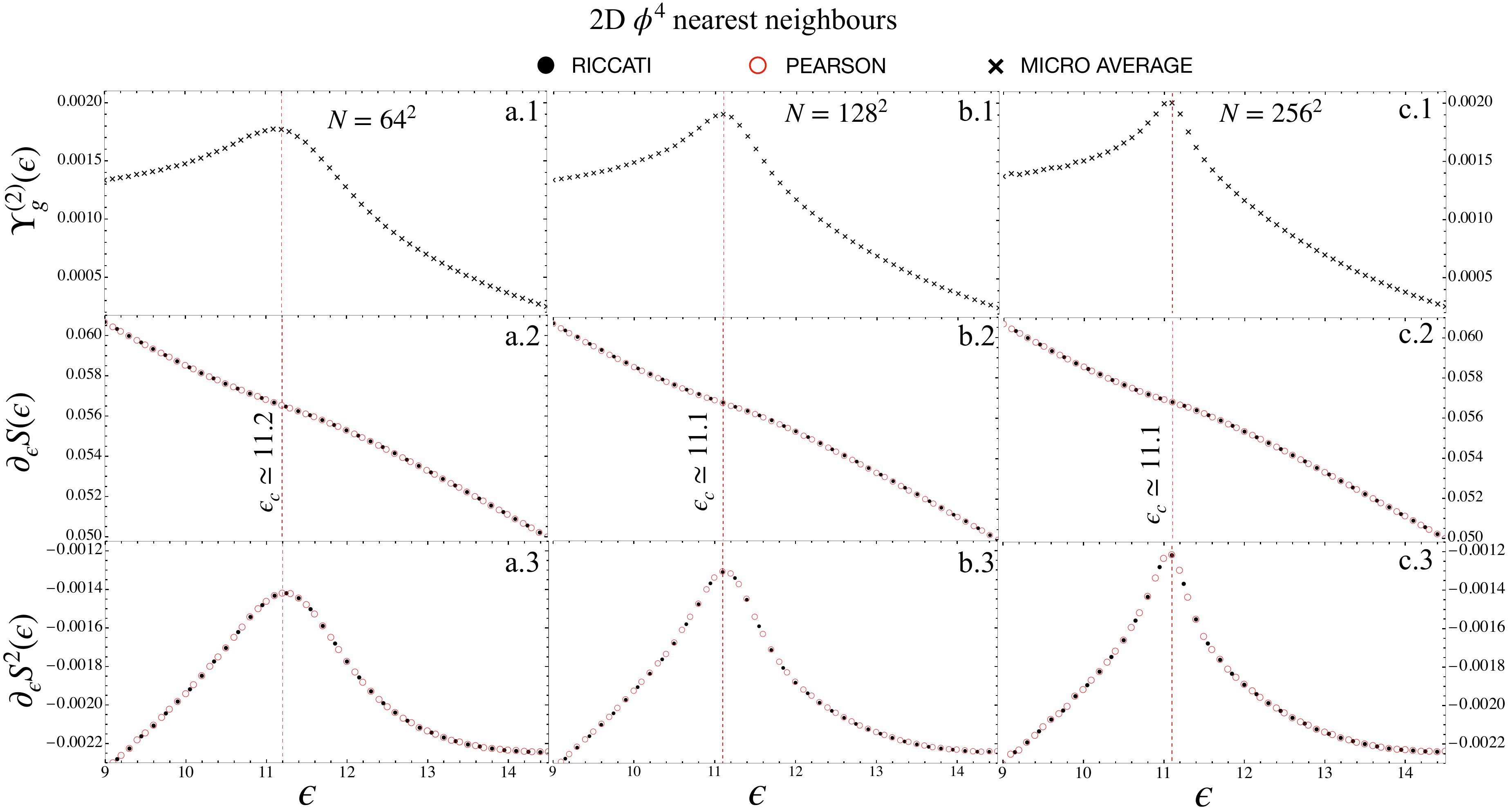}

    \caption{\textbf{Comparison of entropy derivatives at different system sizes obtained as the solution of the entropy flow equation and through thermodynamic methods}. The quantitative and qualitative agreement is excellent. As the system size increases, the peak in $\Upsilon^{(2)}_g$ and $\partial_\epsilon^2 S$ becomes sharper and sharper indicating that in the thermodynamic limit the peak converges to zero thus giving rise to the divergence of the specific heat.}
    \label{fig:solution-riccati-phi}
\end{figure*}

In both cases, we compute the energy derivatives of entropy using the PHT method explained in Sec.~\ref{app:PHT_method}. Notice that the PHT is our reference for the microcanonical observables. This estimation is actually non-perturbative and exact (within statistical precision). In order words, the PHT method expresses thermodynamic derivatives as exact averages of microscopic quantities without approximations, model assumptions, or fitting parameters. Then, we also solved the EFE equation following the procedure of Sec.~\ref{sec:efe-solution-procedure}. Then, for each system size, we computed $\beta^0_N(E_{\text{min}})$ and use it as an initial condition for the differential equation.
This analysis is reported, respectively, in Fig.~\ref{fig:solution-riccati-phi} and~\ref{fig:solution-riccati-xy}.

The solutions of the EFE (black disks) are then plotted in comparison with the PHT-estimation (red circles) for different systems sizes. 

For the $\phi^4$-model, we see a peak (negative-valued maximum) in $\partial_\epsilon^2S$ together with an inflection point in $\partial_\epsilon S$ at $\epsilon_c=11.2$ for $N=64^2$ and $\epsilon_c=11.1$ for $N=128^2$ and $N=256^2$. Moreover, this peak increases with increasing system sizes and approaches zero. According to MIPA \cite{qi2018classification}, this behavior is in agreement with the second-order PT expected in the thermodynamic limit for this system. Moreover, the specific heat
\[
    C_v(E)=\frac{(\partial_E S)^2}{\partial_E^2S}\,
\]
must diverge in the thermodynamic limit (consistently with the definition of second-order transition). This divergence originates from this mechanism. Indeed, $\partial_ES$ manifests a stable inflection point (it does not move for increasing size) while the peak in $\partial_E^2S$ does. This originates the transition.

\begin{figure*}
    \centering
    \includegraphics[width=0.9\linewidth]{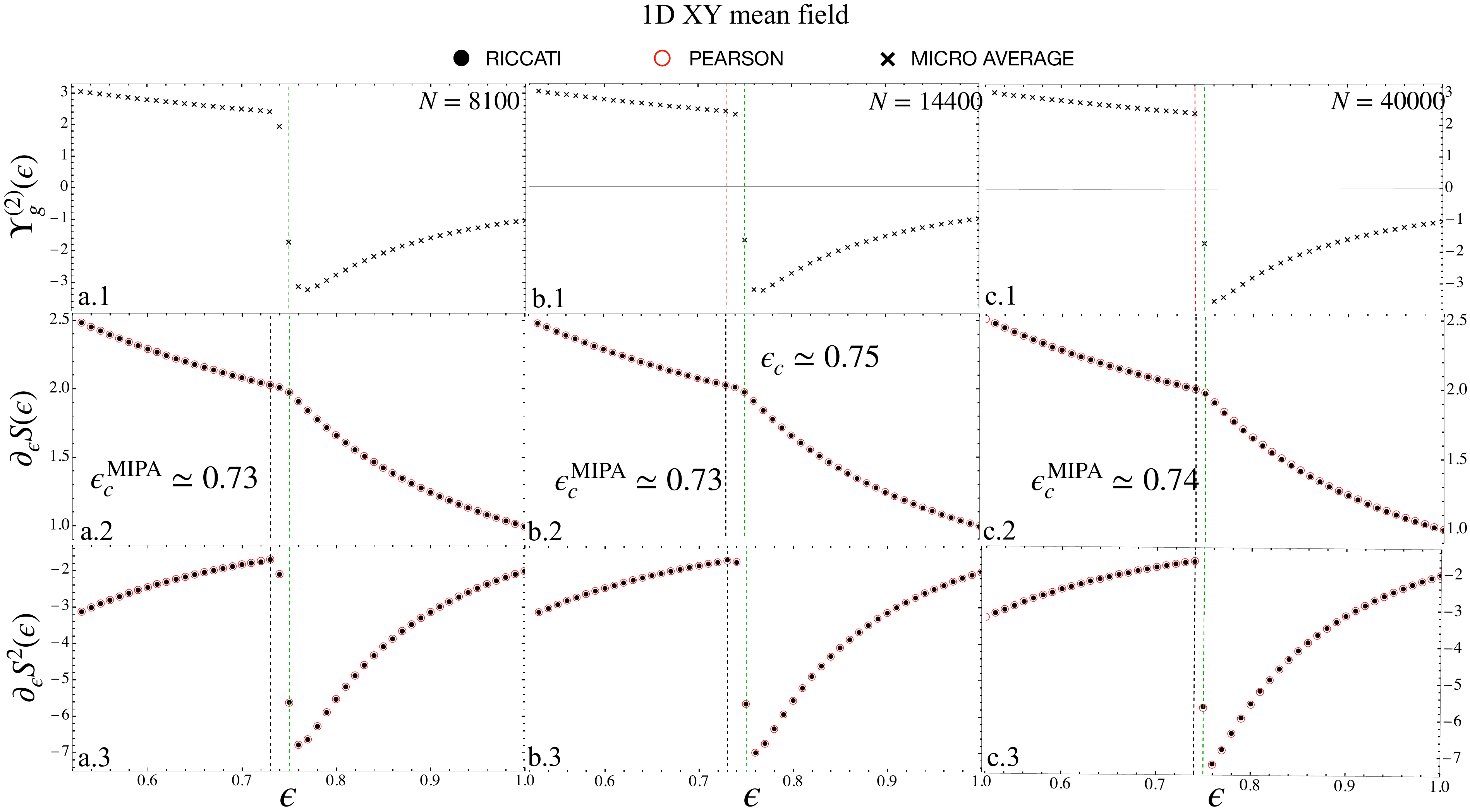}

    \caption{\textbf{Comparison of entropy derivatives at different system sizes obtained as the solution of the entropy flow equation and through thermodynamic methods}. The quantitative and qualitative agreement is excellent also in the mean-field case. Here, the transition is sharper and the discontinuity is visible already at finite size. The geometric function $\Upsilon^{(2)}_g$ and the second derivative $\partial_\epsilon^2 S$ show a jump around $\epsilon_{c}\approx0.74$ closed to the infinite-size energy value $\epsilon_c^\infty=3J/4=0.75$ for our choice $J=1$. Notice that the larger the size, the closer the critical point detected by MIPA.}
    \label{fig:solution-riccati-xy}
\end{figure*}

For the XY mean-field model, MIPA analysis works as well but the mechanism is different. This transition is related to a symmetry breaking associated to the magnetization order parameter and it manifests quite strongly already at finite size. The function $\partial_\epsilon^2S$ admits a negative-valued maximum in $\partial_\epsilon^2S$ but a visible jump appears around the infinity-size critical energy $\epsilon^\infty_c=3J/4=0.75$. The first-derivative $\partial_\epsilon S$ admits a cuspid-like behavior (a drastic inflection point) at $\epsilon_c=0.73$ for $N=8100$ and $\epsilon_c=0.74$ for $N=14400$ and $N=40000$. Interestingly, this maximum converges to the vertical green dashed line corresponding to $\epsilon^\infty_c$ for increasing size and the jump becomes a stronger and stronger discontinuity. This analysis will be reported more deeply in Ref.~\cite{PRE-loris}. The finite-size MIPA analysis is in agreement with the infinite-size PT which is thus classified as a second-order PT.

\end{document}